\documentclass[12pt]{iopart}

\usepackage{graphicx}

\begin{document}

\title{Gravitating Global k-monopole}

\author{Xing-hua Jin, Xin-zhou Li\dag and Dao-jun Liu}

\address{Shanghai United Center for Astrophysics(SUCA), Shanghai
Normal University, Shanghai 200234, China\\
School of Science, East China University of Science and Technology,
130 Meilong Road, Shanghai, 200237, China}

\ead{\dag kychz@shnu.edu.cn}

\begin{abstract}

A gravitating global k-monopole produces a tiny gravitational field
outside the core in addition to a solid angular deficit in the
k-field theory. As a new feature, the gravitational field can be
attractive or repulsive depending on the non-canonical kinetic term.

\end{abstract}

\maketitle

\section{Introduction}

The phase transition in the early universe could produce different
kinds of topological defects which have some important implications
in cosmology\cite{vilenkin}. Domain walls are two-dimensional
defects, and strings are one-dimensional defects. Point-like defects
also arise in same theories which undergo the spontaneous symmetry
breaking, and they appears as monopoles. The global monopole, which
has divergent mass in flat spacetime, is one of the most interesting
defects. The idea that monopoles ought to exist has been proved to
be remarkably durable. Barriola and Vilenkin \cite{barriola} firstly
researched the characteristic of global monopole in curved
spacetime, or equivalently, its gravitational effects. When one
considers the gravity, the linearly divergent mass of global
monopole has an effect analogous to that of a deficit solid angle
plus that of a tiny mass at the origin. Harari and Loust\`{o}
\cite{harari}, and Shi and Li \cite{li1} have shown that this small
gravitational potential is actually repulsive. Furthermore, Li {\it
et al} \cite{li2, li3,li4} have proposed a new class of cold stars
which are called D-stars (defect stars). One of the most important
features of such stars, comparing to Q-stars, is that the theory has
monopole solutions when the matter field is absent, which makes the
D-stars behave very differently from the Q-stars. The topological
defects are also investigated in the Friedmann-Robertson-Walker
spacetime \cite{Basu}. It is shown that the properties of global
monopoles in asymptotically dS/AdS spacetime \cite{li5} and the
Brans-Dicke theory \cite{li6} are very different from those of
ordinary ones. The similar issue for the gravitational mass of
composite monopole, i.e., global and local monopole has also been
discussed \cite{Bezerra}.

The huge attractive force between global monopole $M$ and
antimonopole $\bar{M}$ proposes that the monopole over-production
problem does not exist, because the pair annihilation is very
efficient. Barriola and Vilenkin have shown that the radiative
lifetime of the pair is very short as they lose energy by Goldstone
boson radiation \cite{barriola}. No serious attempt has made to
develop an analytical model of the cosmological evolution of a
global monopole, so we are limited to the numerical simulations of
evolution by Bennett and Rhie \cite{bennett}. In the $\sigma$-model
approximation, the average number of monopoles per horizon is
$N_H\sim4$. The gravitational field of global monopoles can lead to
clustering in matter, and later evolve into galaxies and clusters.
The scale-invariant spectrum of fluctuations has been given
\cite{bennett}. Furthermore, one can numerically obtain the
microwave background anisotropy $(\delta T/T)_{\rm rms}$ patterns
\cite{bennett1}. Comparing theoretical value to the observed rms
fluctuation, one can find the constraint of parameters in global
monopole.

On the other hand, non-canonical kinetic terms are rather ordinary
for effective field theories. The k-field theory, in which the
non-canonical kinetic terms are introduced in the Lagrangian,
 have been recently investigated to serve as the inflaton in the
 inflation scenario, which is so-called k-inflation \cite{k-inflation}, and to explain the current
 acceleration of the universe and the cosmic coincidence
problem, k-essence \cite{k-essence}. Armendariz-Picon {\it et al}
\cite{Halo1,Halo2} have discussed gravitationally bound static and
spherically symmetric configurations of k-essence fields. Another
interesting application of k-fields is topological defects, dubbed
by k-defects \cite{k-defects}. Monopole \cite{li7} and vortex
\cite{liu} of tachyon field, which as an example of k-field comes
from string/M theory, have also been investigated. The mass of
global k-monopole diverges in flat spacetime, just as that of
standard global monopole, therefore, it is of more physical
significance to consider the gravitational effects of global
k-monopole.

In this paper, we study the gravitational field of global k-monopole
and derive the solutions numerically and asymptotically. We find
that the topological condition of vacuum manifold for the formation
of a k-monopole is identical to that of an ordinary monopole, but
their physical properties are disparate. Especially, we show that
the mass of k-monopole can be positive in some form of the
non-canonical kinetic terms. In other words, the gravitational field
can be attractive or repulsive depending on the non-canonical
kinetic term.

\section{Equations of Motion}

We shall work within a particular model in units $c=1$, where a
global $O(3)$ symmetry is broken down to $U(1)$ in the k-field
theory. Its action is given by
\begin{equation}\label{S0}
S=\frac{1}{2\kappa}\int
d^4\tilde{x}\sqrt{-g}\left[M^4K(\tilde{X}/M^4) -
\frac{1}{4}\lambda^2(\tilde{\phi}^a\tilde{\phi}^a-\tilde{\sigma}_0^2)^2\right],
\end{equation}
where $\kappa=8\pi G$ and $\lambda$ is a dimensionless constant. In
action (\ref{S0}),
$\tilde{X}=\frac{1}{2}\left(\tilde{\partial}_{\mu}\tilde{\phi}^a\tilde{\partial}^{\mu}\tilde{\phi}^a\right)$,
where $\tilde{\phi}^a$ is the $SO(3)$ triplet of goldstone field and
$\tilde{\sigma}_0$ is the symmetry breaking scale with a dimension
of mass. After setting the dimensionless quantities: $x=M\tilde{x}$,
$\phi_a=\tilde{\phi}_a/M$ and $\sigma_0=\tilde{\sigma}_0/M$, action
(\ref{S0}) becomes
\begin{equation}\label{S}
S=\frac{1}{2\kappa}\int d^4x\sqrt{-g}\left[K(X) -V(\phi)\right],
\end{equation}
where $V(\phi)=\frac{1}{4}\lambda^2(\phi^a\phi^a-\sigma_0^2)^2$. The
hedgedog configuration describing a global k-monopole is
\begin{equation}\label{config}
  \phi^a=\sigma_0f(\rho)\frac{x^a}{\rho},
\end{equation}
where $x^ax^a=\rho^2$ and $a=1, 2, 3$, so that we shall actually
have a global k-monopole solution if $f\rightarrow 1$ at spatial
infinity and $f\rightarrow0$ near the origin.

The static spherically symmetric metric can be written as
\begin{equation}\label{metric}
  ds^2=B(\rho)dt^2-A(\rho)d \rho^2-\rho^2(d\theta^2+ \sin^2\theta
  d\varphi^2)
\end{equation}
with the usual relation between the spherical coordinates $\rho$,
$\theta$, $\varphi$ and the "Cartesian" coordinate $x^a$.
Introducing a dimensionless parameter $r=\sigma_0 \rho$, from
(\ref{S}) and (\ref{config}), we obtain the equations of motion for
$f$ as
\begin{eqnarray}\label{gold}
\dot{K}\left\{\frac{1}{A}f^{\prime\prime} +
\left[\frac{2}{Ar}+\frac{1}{2B}\left(\frac{B}{A}\right)'\right]f'-\frac{2}{r^2}f\right\}+
\ddot{K}\frac{X'f'}{A} - \lambda^2f(f^2-1)=0,
\end{eqnarray}
where the prime denotes the derivative with respect to $r$, the dot
denotes the derivative with respect to $X$ and
$X=-\frac{f^2}{r^2}-\frac{f'^2}{2A}$. Since we only consider the
static solution, positive $X$ and negative $X$ are irrelevant each
other. In this paper, we will assume $K(X)$ to be valid for negative
$X$.

The Einstein equation for k-monopole is
\begin{equation}\label{einsteineq}
G_{\mu\nu}=\kappa T_{\mu\nu}
\end{equation}
where $T_{\mu\nu}$ is the energy-momentum tensor for the action
(\ref{S}). The tt and rr components of the Einstein equations now
could be written as
\begin{eqnarray}\label{e00}
-\frac{1}{A}\left(\frac{1}{r^2}-\frac{1}{r}\frac{A'}{A}\right)+\frac{1}{r^2}&=&\epsilon^2T^0_0\label{e01}\\
-\frac{1}{A}\left(\frac{1}{r^2}+\frac{1}{r}\frac{B'}{B}\right)+\frac{1}{r^2}&=&\epsilon^2T^1_1\label{e11},
\end{eqnarray}
where
\begin{eqnarray}\label{t00}
T^0_0&=&-K+\frac{\lambda^2}{4}(f^2-1)^2\\
T^1_1&=&-K+\frac{\lambda^2}{4}(f^2-1)^2
-\dot{K}\frac{f'^2}{A}\label{t11}
\end{eqnarray}
and $\epsilon^2=\kappa\sigma_0^2=8\pi G\sigma_0^2$ is a
dimensionless parameter.

\section{k-monopole}

Although the existence of global k-monopole, as well as the standard
one, is guaranteed by the symmetry-breaking potential, there exist
the non-canonical kinetic term in k-monopole which certainly leads
to the appearance of a new scale in the action and the mass
parameter in the potential term. However, the non-canonical kinetic
term is non-trivial. At small gradients, it can be chosen to have a
same asymptotical behavior with that of the standard one, so that it
ensures the standard manner of a small perturbations. While at large
gradient we choose it to have a different form from the standard
one.

In the small $X$ case, we assume that the kinetic term has the
asymptotically canonical behavior, which can avoid "zero-kinetic
problem". If $|X|\ll 1$, we have $K(X)\sim X^{\alpha}$, and
$\alpha<1$ then there is a singularity at $X=0$; and $\alpha >1$
then the system becomes non-dynamical at $X=0$. For the monopole
solution, it is easily found that $K(X)\sim X$ at $r\gg 1$. On the
other hand, we assume that the modificatory kinetic term $K(X)\sim
X^{\alpha}$ and $\alpha\neq 1$ at $|X|\gg 1$. One can easily obtain
the equation of motion inside the core of a global monopole after
assuming that $|X|\gg 1$ in the core of the global monopole. The
equations of motion are highly non-linear and cannot be solved
analytically. Next, we investigate the asymptotic behaviors of
global monopole with non-linear in $X$ kinetic term. To be specific,
we consider the following type of kinetic term
\begin{equation}\label{kx}
K(X)=X-\beta X^2,
\end{equation}
where $\beta$ is a parameter of global k-monopole. It is easy to
find that global k-monopole will reduce to be the standard one when
$\beta=0$. It is easy to check whether the kinetic term (\ref{kx})
satisfy the condition for the hyperbolicity \cite{Halo2, Rendall,
k-defects}
\begin{equation}
  \label{hyperbolicity}
  \frac{\dot{K}}{2X \ddot{K}+\dot{K}}>0,
\end{equation}
which leads to a positive definite speed of sound for the small
perturbations of the field. The stability of solutions shows that
for the case $\beta<0$, the range
$\frac{1}{6\beta}>X>\frac{1}{2\beta}$ must be excluded. However,
this will not destroy the results which are carried out from the
case $\beta>0$. We here only consider the cases for $\beta>0$.

Using Eqs.(\ref{gold})-(\ref{t11}), we get the asymptotic expression
for $A(r)$, $B(r)$, and $f(r)$ which is valid near $r=0$,
\begin{eqnarray}\label{fs}
f(r)&=&f_0{r}+\frac{{f_0}\,\left[ 2\,\lambda^2 \,\left( -3 +
\epsilon^2 \right)+
 +42\beta\,\epsilon^2 \,{{f_0}}^4
\right]}{60\,\left( 1 + 5\beta\,{{f_0}}^2 \right)}{r}^3
\\\nonumber
&+& \frac{f_0\left[\left( 9 + 7\beta\,\lambda^2 \right) \,\epsilon^2
\,{{f_0}}^2+ 36\beta^2\,\epsilon^2 \,{{f_0}}^6 \right] }{60\,\left(
1 + 5\beta\,{{f_0}}^2 \right)
      }{r}^3+O(r^4)
\end{eqnarray}
\begin{equation}
A(r)=1+\frac{\epsilon^2 \,\left( \lambda^2  + 6\,{{f_0}}^2 +
9\beta\,{{f_0}}^4 \right) }{12}{r}^2+O(r^3)
\end{equation}
\begin{equation}
B(r)=1+\frac{\epsilon^2 \,\left( \lambda^2  - 9\beta\,{{f_0}}^4
\right) }{12}{r}^2+O(r^3),
\end{equation}
where the undetermined coefficient $f_0$ is characterized as the
mass of k-monopole, which can be determined in the numerical
calculation.

In the region of $r\gg 1$, similarly we can expand $f(r)$, $A(r)$
and $B(r)$ as
\begin{eqnarray}\label{fb}
f(r)&=&1-\frac{1}{\lambda^2}\left(\frac{1}{r}\right)^2-\frac{3 -
2\,\epsilon^2 + 4\beta\,\lambda^2}
  {2\,{\lambda^4 }}\left(\frac{1}{r}\right)^4+O(r^{-5})
\\
A(r)&=&\frac{1}{1-\epsilon^2}-\frac{{M_{\infty}}}{{(1-\epsilon^2)^2}}\frac{1}{r}+\left[\frac{\epsilon^2(1
- \beta\lambda^2) }
  {{\left( 1 - \epsilon^2  \right) }^2\,\lambda^2 }+\frac{{M_{\infty}^2}}{{(1-\epsilon^2)^3}}\right]
  \left(\frac{1}{r}\right)^2\\\nonumber
  &+&O(r^{-3})
\\\label{br}
B(r)&=&(1-\epsilon^2)+M_{\infty}\frac{1}{r} - \frac{\epsilon^2
\,\left( 1 -\beta \lambda^2 \right)}
    {\lambda^2 } \left(\frac{1}{r}\right)^2+O(r^{-3}),
\end{eqnarray}
where the constant $M_{\infty}$ will be discussed in the following.

\begin{figure}
\begin{center}
\includegraphics[width=3in]{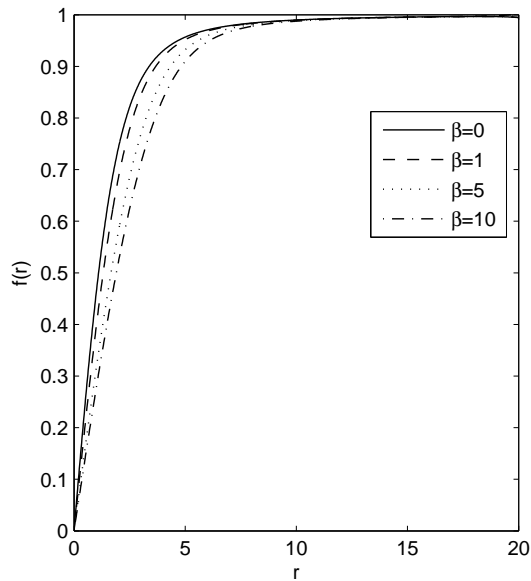}
\caption{The plot of $f(r)$ as a function of $r$. Here we choose
$\lambda = 1$, $G=1$ and $\epsilon=0.001$. The four curved lines are
plotted when $\beta=0$, $\beta=1$, $\beta=5$ and $\beta=10$
respectively. } \label{f}
\end{center}
\end{figure}

Using shooting method for boundary value problems, we get the
numerical results of the function $f(r)$ which describes the
configuration of global k-monopole. In Fig.\ref{f} we show the
function $f(r)$ for $\beta=0$, $\beta=1$, $\beta=5$ and $\beta=10$
respectively and for given values of $\lambda$ and $\epsilon$.
Obviously, the configuration of field $f$ is not impressible to the
choice of the parameter $\beta$.

From Eqs.(\ref{fs}) and (\ref{fb}), it is easy to construct global
k-monopole which has the same asymptotic condition with standard
global monopole, \emph{i.e.}, $f$ will approach to zero when $r \ll
1$ and approach to unity when $r \gg 1$.

Actually there is a general solution to Einstein equation with
energy-momentum tensor $T_{\mu\nu}$ which takes the form as
(\ref{t00}) and (\ref{t11}) for spherically symmetric metric
(\ref{metric})
\begin{equation}
\label{19}
A(r)^{-1}=1-\frac{\epsilon^2}{r}\int_0^r[-K+\frac{\lambda^2}{4}(f^2-1)^2]r^2dr
\end{equation}
\begin{equation}
B(r)=A(r)^{-1}\exp\left[\epsilon^2\int_{\infty}^r(\dot{K}{f'^2}r)dr\right].
\end{equation}
In terms of the dimensionless quantity $\epsilon$, the metric
coefficient $A(r)$ and $B(r)$ can be formally integrated and read as
\begin{equation}\label{solution1}
 A(r)^{-1}=1-\epsilon^2-\frac{2G\sigma_0M_A(r)}{r}
\end{equation}
\begin{equation}\label{solution2}
 B(r)=1-\epsilon^2-\frac{2G\sigma_0M_B(r)}{r}.
\end{equation}
The small dimensionless parameter $\epsilon$ arise naturally from
Einstein equations and clearly $\epsilon^2$ describes a solid
angular deficit of space-time.

A global k-monopole solution $f$ should approach unity when $r\gg1$.
If this convergence is fast enough, then $M_A(r)$ and $M_B(r)$ will
also rapidly converge to finite values. Therefore, from
Eqs.(\ref{fb})-(\ref{br}) we have the asymptotic expansions:
\begin{eqnarray}
M_A(r)&=&{M_{\infty}}+4\pi \sigma_0 \,\left( -\beta  +
\frac{1}{{\lambda }^{2}} \right)\frac{1}{r} +
  \frac{8\pi \sigma_0\left( -1 + \epsilon^2  + 2\,\beta \,{\lambda }^2 \right)
       }{3\,{\lambda }^4}\left(\frac{1}{r}\right)^3
       \\\nonumber &+&O(r^{-5})\\
M_B(r)&=&{M_{\infty}}+4\pi \sigma_0 \,\left( -\beta  +
\frac{1}{{\lambda }^{2}} \right)\frac{1}{r} -
  \frac{4\pi \sigma_0\,\left( -1 + \epsilon^2  - 4\,\beta \,{\lambda }^2 \right) }
   {3\,{\lambda }^4}\left(\frac{1}{r}\right)^3
   \\\nonumber &+&O(r^{-5}),
\end{eqnarray}
where $M_{\infty}\equiv \lim _{r\rightarrow \infty}M_A(r)$. One can
easily find that the dependence on $\epsilon$ of the asymptotic
expansion for $f(r)$ is very weak, in other words, the asymptotic
behavior is quite independent of the scale of symmetry breakdown
$\sigma_0$ up to value as large as the planck scale. On the
contrary, $M_A(r)$ depends obviously on $\sigma_0$.

\begin{figure}
\begin{center}
\includegraphics[width=3in]{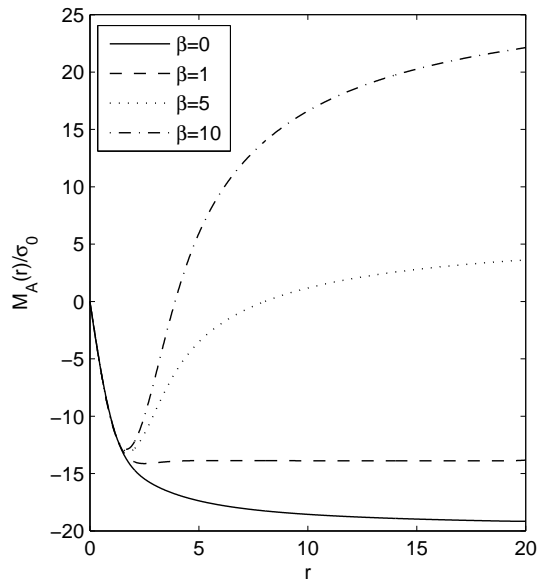}
\caption{The plot of $M_A(r)/\sigma_0$ as a function of $r$. Here we
choose $\lambda = 1$, $G=1$ and $\epsilon=0.001$. The four curved
lines are plotted when $\beta=0$, $\beta=1$, $\beta=5$ and
$\beta=10$ respectively.} \label{ma}
\end{center}
\end{figure}

The numerical results of $M_A(r)/\sigma_0$ are shown in Fig.\ref{ma}
by shooting method for boundary value problems where we choose
$\lambda = 1$, $G=1$ and $\epsilon=0.001$. From the figure, we find
that the mass of global k-monopole decrease to a negative asymptotic
value when $r$ approaches infinity in the case that $\beta=0$ and
$\beta=1$. While the mass will be positive if $\beta=5$ or
$\beta=10$. The asymptotic mass for the cases above are
$-19.15\sigma_0$, $-13.83\sigma_0$, $3.62\sigma_0$ and
$22.14\sigma_0$ respectively. It is clear that the presence of
parameter $\beta$, which measures the degree of deviation of kinetic
term from canonical one, affects the effective mass of the global
k-monopole significantly. It is not difficult to understand this
property. From Eqs.(\ref{kx}), (\ref{19}) and (\ref{solution1}), the
mass function $M_A(r)$ can be expressed explicitly as
\begin{equation}
 \frac{M_A(r)}{4\pi\sigma_0}=-r+\int_0^r\left[\beta X^2-X+\frac{\lambda^2}{4}(f^2-1)^2\right]r^2dr.
\end{equation}
Obviously, the $\beta$-term in the integration has the positive
contribution for the mass function. From Fig.\ref{f}, $f$ (then $X$)
is not sensitive to the value of $\beta$, so the greater parameter
$\beta$ is chosen, the larger value $M_A(r)$ takes for a given $r$,
and if $\beta$ is greater than some value, $M_A(r)$ will be positive
for  large $r$ as Fig.\ref{ma} shows. However, inside the core, $X$
varies slowly but $f$ varies fast with respect to $r$, therefore,
$\lambda^2$-term in the integration will become dominant as $r$
decreases. This leads to two characteristics of the mass curves
which is also shown in Fig.\ref{ma}: (i) the mass curves with
different $\beta$ converge gradually in the region near $r=0$; (ii)
in the case that $\beta$ is large enough, $M_A(r)$ has a minimum.

To show the effect of solid defect angle, we then investigate the
motion test particles around a global k-monopole. It is a good
approximation to take $M_A(r)$ as a constant in the region far away
from the core of the global k-monopole, since the the mass $M_A(r)$
approaches very quickly to its asymptotic value. Therefore, we can
consider the geodesic equation in the metric (\ref{metric}) with
\begin{equation}\label{solution22}
 A(r)^{-1}=B(r)=1-\epsilon^2-\frac{2G M}{r}
\end{equation}
where $M=\sigma_0M_{\infty}$. Solving the geodesic equation and
introducing a dimensionless quantaty $u=GM/r$, one will get the
second order differentiating equation of $u$ with respect to
$\varphi$ \cite{li5, wald}
\begin{equation}
\frac{d^2u}{d\varphi^2}+(1-\epsilon^2)u=\left(\frac{GM}{L}\right)^2+3u^2,
\end{equation}
where $L$ is the angular momentum per unit of mass. When
$\left(\frac{GM}{L}\right)^2\varphi\ll 1$, one have the approximate
solution of $u$
\begin{equation}
u\approx\left(\frac{GM}{L}\right)^2\left\{\frac{1}{1-\epsilon^2}
+e\cos\left[\left(1-\frac{3}{\sqrt{\left(1-\epsilon^2 \right)^3}}
\left(\frac{GM}{L}\right)^2 \right)\varphi\right]\right\},
\end{equation}
where $e$ denotes the eccentricity. When a test particle rotates one
loop around the global k-monopole, the precession of it will be
\begin{equation}\label{d}
\Delta\varphi=6\pi\left(\frac{GM}{L}\right)^2\frac{1}{\sqrt{\left(1-\epsilon^2
\right)^3}}\approx
6\pi\left(\frac{GM}{L}\right)^2+9\pi\left(\frac{GM}{L}\right)^2\epsilon^2.
\end{equation}
The last term in Eq.(\ref{d}) is the modificaiton comparing this
result with that for the precession around an ordinary star.

\section{Conclusion}
In summary, k-monopole could arise during the phase transition in
the early universe. We calculate the asymptotic solutions of global
k-monopole in static spherically symmetric spacetime, and find that
the behavior of a k-monopole is similar to that of a standard one.
Although the choice of the parameter $\beta$, which measures the
degree of deviation of kinetic term from canonical one, have little
influence on the configuration of k-field $\phi$, the effective mass
of global k-monopole is affected significantly. The mass might be
negative or positive when different parameters $\beta$ are chosen.
This shows that the gravitational field of the global k-monopole
could be attractive or repulsive depending on the different
non-canonical kinetic term.

The configuration of a global k-monopole is more complicated than
that of a standard one. As for its cosmological evolution, we should
not attempt to get the analytical mode, instead we can only use
numerical simulation. However, the energy dominance of global
k-monopole is in the region outside the core. We can roughly
estimate that global k-monopoles will result in the clustering in
matter and evolve into galaxies and clusters in a way similar to
that of standard monopoles.
\section*{Acknowledgement}

This work is supported in part by National Natural Science
Foundation of China under Grant No. 10473007 and No. 10503002 and
Shanghai Commission of Science and Technology under Grant No.
06QA14039.

\end{document}